\documentclass[float,preprintnumbers,amssymb,twocolumn,superscriptaddress,aps,prl]{revtex4-1}
\usepackage{graphicx}
\usepackage{booktabs}
\usepackage{amsfonts, amsmath, amsthm, amssymb,mathrsfs} % For math f
\usepackage{mathtools}
\usepackage{wrapfig, tikz,url,lipsum,float}
\usepackage[normalem]{ulem}
\usepackage{dcolumn}
\usepackage{bm}
\usepackage{color}
\usepackage[colorlinks=true,linkcolor=blue,citecolor=blue]{hyperref}
\usepackage{hyperref}
\usepackage{footnote}
\usepackage{cleveref}

\newcommand{\REM}[1]{{}}
\newcommand{\dx}[1]{\frac{\partial{#1}}{\partial x}}
\newcommand{\dxx}[1]{\frac{\partial^2{#1}}{\partial x^2}}
\newcolumntype{L}{>{\centering\arraybackslash}m{3cm}}

\begin{document}

\date{Revised \today}
\title{\bf Fixation in Competing Populations: Diffusion and Strategies for Survival}

\author{Tapas Singha} 
%\email{tapass@tifrh.res.in}
\author{Prasad Perlekar}
%\email{barma23@gmail.com}
\author{Mustansir Barma}
\affiliation{TIFR Centre for Interdisciplinary Sciences, Tata Institute of Fundamental Research, Gopanpally, Hyderabad 500107, India}

\begin{abstract}
How should dispersal strategies be chosen to increase the likelihood of survival of a species? We obtain the answer for the spatially extended versions of three well-known models of two competing species with unequal diffusivities. Though identical at the mean-field level, the three models exhibit drastically different behaviour leading to different optimal strategies for survival, with or without a selective advantage for one species. With conserved total particle number, dispersal has no effect on survival probability. With a fluctuating number, faster dispersal is advantageous if intra-species competition is present, while moving slower is the optimal strategy for the disadvantaged species if there is no intra-species competition: it is imperative to include fluctuations to properly formulate survival strategies.
\end{abstract}

\maketitle

\section{Introduction}
Biological dispersal refers to the movement of individuals and is a key feature of population dynamics. Dispersal has consequences not only for species distribution but also for individual survival, thereby influencing many different aspects of evolutionary dynamics, from epidemic outbreaks to the evolution of language \cite{Smith1982, Hofbauer1998, Nowak2006, ColizzaPRL2007,Baxter2008, Hamilton2009,Levin1974}. The importance of dispersal rates for ultimate survival stems from the fact that two individuals need to be in the same locality in order to interact. Evidently, the nature of interactions is also crucial, both within species and across species, possibly including selective advantage that favours one species. Thus the likelihood that a particular species eventually prevails depends on multiple factors \cite{Murray1, Murray2, Strogatz,Hartl89,Korolev2010,Frey07}. Indeed, when intra-species interactions dominate, the optimal strategy for choosing dispersal rates has been explored earlier. In the presence of logistic-type \emph{competition} for resources between members of the same species, the survival probability of a species increases as the dispersal rate increases \cite{Pigolotti2015,Pigolotti2016,Korolev2019}, while in the presence of \emph{cooperation} between members of the same species, the survival probability increases as the dispersal rate decreases \cite{Korolev2015}. In general, how should dispersal strategies be chosen so as to increase the likelihood of survival of a species?

In this paper, we address this broad and important question by focusing on the dispersal properties of  two competing species in a spatially extended system. The dynamics involves random diffusive motion and may also include stochastic birth and death of competing individuals. Indeed, scenarios where the number of individuals is not fixed are ubiquitous \cite{Murray1,Murray2,Strogatz,Korolev2010}.  We will see below that choosing an optimal strategy to maximize survival probability needs a nuanced understanding of factors which arise from the dynamics.

To understand these factors, we carry out a parallel investigation of three simple well-studied models with different reaction dynamics \cite{Bhat2019, pigo13, doe03, Chotibut2015, Rana2017}. We will show that, although for equal-diffusivity cases with no selective advantage, all the models have similar statistical behavior, a slight imbalance in either the diffusivity or the selective advantage dramatically alters the outcome.

\section{Models}
The models are defined on a one-dimensional (1D) lattice of $L$ collocated points with unit separation and involve two competing species (say $A$ and $B$). Interactions are local and involve two individuals at a time on the same lattice site. We start with an initially well-mixed population with a number $N/2$ of $A$ and $B$ particles. The onsite interactions in the three models of interest are defined by Eqs.\,\eqref{Reaction_VM}$-$\eqref{Reaction_CLVM} below; in every case, these are supplemented by diffusive dispersal.

\emph{Voter-type Model with Diffusion} (VMD): In a single microstep, the inter-species reactions on each site follow Moran dynamics with rates $\lambda(1+\frac{1}{2}s)$ and $\lambda(1-\frac{1}{2}s)$:
\begin{equation}
A+B \xrightarrow {\lambda(1+\frac{1}{2}s)} A+A; \, \, \, \, \, \, \, B+A \xrightarrow {\lambda(1-\frac{1}{2}s)} B+B
\label{Reaction_VM}
\end{equation}
Here $s$ is the selective advantage, which gives a preference to either $A (s >0)$ or $B (s < 0)$ in the competition, with $s=0$ being the neutral case. For $s>0$ ($s<0$), species $A$ ($B$) has a selective advantage.

Evidently the total number of particles in the full system is strictly conserved, but on each site the evolution differs from the strict Moran process \cite{Moran58,Crow70,Hartl89,Korolev2010} as the number of particles fluctuates owing to diffusion. The dynamics on each site  resembles that of the voter model \cite{Clifford1973, Holley1975}.\REM{ which has been used as a model of evolution of opinion \cite{Jedrzejewski2019,Blythe2007}, and disease spreading \cite{Pinto2011}.}

\emph{Fluctuating Voter-type Model with diffusion} (FVMD): In addition to the competitive Moran moves Eq. \eqref{Reaction_VM}, we allow individuals to give birth or die, at equal rates $\mu$ \cite{Bhat2019}. 
\begin{eqnarray}
A \xrightarrow {\mu}&& 2A; \, \, \, A \xrightarrow {\mu} 0; \,  \, \, A+B \xrightarrow {\lambda(1+\frac{1}{2}s)} A+A;    \nonumber \\
B \xrightarrow {\mu} && 2B; \, \, \, B \xrightarrow {\mu} 0;  \, \, \, B+A \xrightarrow {\lambda(1-\frac{1}{2}s)} B+B 
\label{Reaction_FVM}
\end{eqnarray}

\onecolumngrid

\begin{figure}[H]
%\centering
%\hspace{1.38cm}
%\includegraphics[width=0.68 \linewidth, height=0.065\linewidth]{Names.pdf}\\
\includegraphics[width=1.0\linewidth, height=0.68\linewidth]{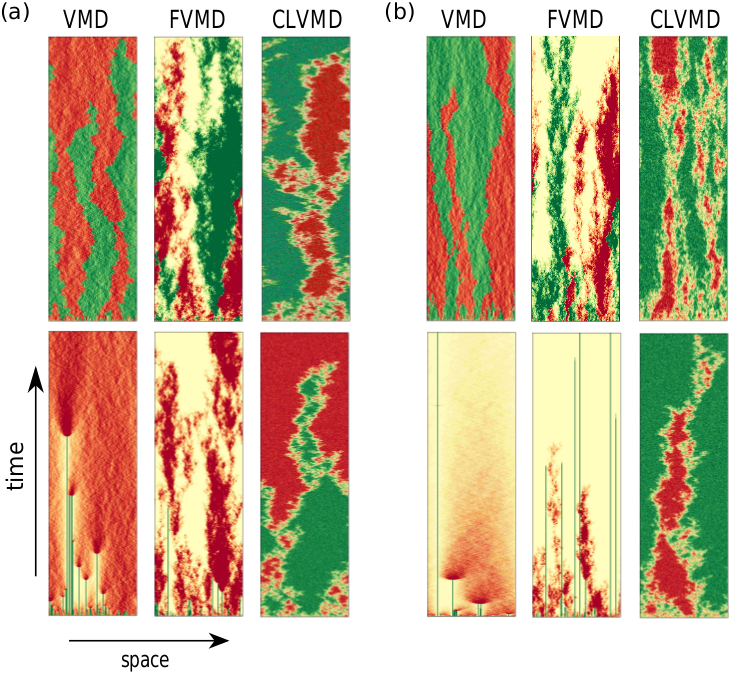}
\caption{\label{World_Lines} Space-time plots for the three models, namely Voter Model with Diffusion (VMD), Fluctuating Voter Model with Diffusion (FVMD), and Competitive Lotka-Volterra Model with Diffusion (CLVMD), with equal (top panels) and unequal (bottom panels) diffusivities and with zero selective advantage. Red (species A) and green (species B) indicate two competing species, with green diffusing slower (bottom panels), with the intensity of color being an indicator of local density. Yellow indicates empty space. At left (a), the top and bottom panels are shown when $A$ wins, and similarly, at right (b), top and bottom panels are shown when $B$ wins. The initial state has well-mixed populations with equal densities $\rho_A=\rho_B=8$ of the two species, where $\rho_A=N/(2L)$. The diffusion constant for the slower species is taken to be zero for the VMD and FVMD, resulting in pillar-like structures localized at sites. For CLVMD, the slower species has diffusion constant 0.96. It is evident that the dynamics of fluctuations is very different in the three cases. }
\end{figure}

\twocolumngrid

In our numerical work, we choose $\mu=\lambda=2$. The dynamics leads to a quasi-steady state where the total number of particles fluctuates around an average value and one of the two species fixates \cite{gab13}. This is followed by, on very long times, an overall extinction (see Appendix A).

\emph{Competitive Lotka-Volterra Model with Diffusion} (CLVMD): In this model, individuals can give birth at rate $\mu$ but death occurs because of intra-species and inter-species competition, at rates $\gamma_1$ and $\gamma_2(1\pm s)$ respectively  \cite{Pigolotti2015, Chotibut2015}. The reactions are 
\begin{eqnarray}
 A\xrightarrow {\mu}&&2A; \, \,  \,  A+A \xrightarrow {\gamma_1} A; \, \,  \,  A+B \xrightarrow {\gamma_2 (1+s)} A; \nonumber \\ 
 B\xrightarrow {\mu} && 2B; \, \, \, B+B \xrightarrow {\gamma_1} B; \, \, \, B+A \xrightarrow {\gamma_2 (1-s)} B 
\label{Reaction_CLVM}
\end{eqnarray}
In steady state the mean concentration or carrying capacity $\rho$ is given by $\mu/\gamma_1$. 

%%%%%%%%%%%%%%%%%%%%%%%%%%%%%%%%%%%%%%%%%%%%%%%%%%%%%%%%%%%%%%%%%%%
%%%%%%%%%%%%%%%%%%%%%%%%%%%%%%%%%%%%%%%%%%%%%%%%%%%%%%%%%%%%%%%%%%%%

%\begin{figure}[H]
%\begin{center}
%%\hspace{-0.24cm}
%\includegraphics[width=0.6\linewidth, height=0.06\linewidth]{Names.pdf}\\
%\includegraphics[width=0.25\linewidth, height=0.55\linewidth]{VMD_Eq_BWINS.png}
%%\hspace{0.05cm}
%\includegraphics[width=0.25\linewidth, height=0.53\linewidth]{FVMD_Eq_BWINS.png}
%%\hspace{0.05cm}
%\includegraphics[width=0.25\linewidth, height=0.55\linewidth]{CLVMD_Eq_BWINS.png}\\
%\hspace{-1cm}
%\includegraphics[width=0.08\linewidth, height=0.5\linewidth]{Axis_Time.pdf}
%\includegraphics[width=0.25\linewidth, height=0.54\linewidth]{VMD_BWins.jpeg}
%%\hspace{0.05cm}
%\includegraphics[width=0.25\linewidth, height=0.54\linewidth]{FVMD_BWins.jpeg}
%%\hspace{0.05cm}
%\includegraphics[width=0.25\linewidth, height=0.54\linewidth]{CLVMD_BWins.jpeg}\\
%\hspace{0.2cm}
%\includegraphics[width=0.8\linewidth, height=0.08\linewidth]{Axis_Space.pdf}\\
%\end{center}
%\caption{\label{World_Lines1} \textcolor{blue}{Space-time plots for the three models, namely VMD, FVMD, and CLVMD when $B$ wins, with equal diffusivities (top panel) and unequal diffusivities (bottom panel). Yellow indicates empty space. The slower species (shown in green) has diffusion constant zero for the VMD and FVMD, resulting in pillar-like structures localized at sites. For CLVMD, the slower species has diffusion constant $0.96$. It is evident that the dynamics of fluctuations is very different in the three cases.}}
%\end{figure}

In addition to the reactions [Eqs. \eqref{Reaction_VM} - \eqref{Reaction_CLVM}], individuals can move stochastically to neighboring sites but with different hopping rates for $A$ and for $B$, reflecting different dispersal rates $D_A$ and $D_B$.
\REM{%
\begin{eqnarray} 
\hspace{-0.8mm}
\{\cdot \cdot; n^{A}_{i}, n^{B}_{i}; n^{A}_{j}, n^{B}_{j};\cdot && \cdot \}  \xrightarrow {D_A} 
\{ \cdot \cdot; n^{A}_{i}-1, n^{B}_{i}; n^{A}_{j}+1, n^{B}_{j};  \cdot \cdot \} \nonumber \\
&& \xrightarrow {D_B} \{ \cdot \cdot; n^{A}_{i}, n^{B}_{i}-1; n^{A}_{j}, n^{B}_{j}+1; \cdot \cdot \} 
\label{Diff}
\end{eqnarray}
Here $n^{A}_i$, $n^{B}_i(=0,1,2,.,..)$ are the $A$ and $B$ particle occupancies on site $i$ and $j=i\pm 1$ and $n^{A}_i \rightarrow n^{A}_{i-1}$ moves occur only if $n^{A}_i > 0$, with a similar condition for $B$.} Note that unlike the stepping stone model \cite{Kimura1953, KimuraWeiss1964, Korolev2010}, the number of individuals on a site can fluctuate.

The models which are defined by Eqs.\,\eqref{Reaction_VM}$-$\eqref{Reaction_CLVM} show very different behavior, as evidenced by Fig.~\ref{World_Lines}, which shows the time evolution for the three, with equal (top panels) and unequal (bottom panels) dispersal rates and with $s=0$ for the two species. This happens, as we show later, even though the mean field and well-mixed \cite{pigo13} descriptions of all three are identical, underscoring the importance of fluctuations in formulating survival strategies.

\section{Mean field equations}
The mean field equations for the models can be derived from the corresponding master equation \cite{pigo13_1,doe03_1,red13,risken,kampen}.  Below we present the mean field equation for the concentration of the two species ($c_A$ and $c_B$) for each of models and from it derive the equation for the total concentration $c=c_A+c_B$ and the fraction $f=c_A/c$.

\subsection{Mean field equation for VMD and FVMD}
Note that because the birth-rate and death-rate in FVMD are the same, the mean field equation for VMD and FVMD are identical. The equations for the concentration of $A$ and $B$ species are:
\begin{eqnarray}
\frac{d c_A}{dt} &=&  \frac{\lambda }{2} s c_A c_B + D_A \dxx{c_A}, \\
\frac{d c_B}{dt} &=& -\frac{\lambda }{2} s c_A c_B + D_B \dxx{c_B}.
\end{eqnarray}
Using the above equations for $c_A$ and $c_B$, we obtain the following equations $c$ and $f$:
\begin{eqnarray}
\frac{d c}{dt} &=& (D_A-D_B) \dxx{(c f)}+ D_B \dxx{c}, \\
\frac{d f}{dt} &=& \lambda s f (1-f) + D_A \dxx{f} + \frac{2D_A}{c} \dx{c}\dx{f} \nonumber \\ &+& f \frac{D_A-D_B}{c}\dxx{c} - f \frac{D_A-D_B}{c}\dxx{(c f)}.
\label{MFQ_VMD_FVMD}
\end{eqnarray}
For $D_A=D_B$, Eq.\,\eqref{MFQ_VMD_FVMD} reduces to the Fisher equation \cite{Fisher1937,Kolmogorov1937}.

\subsection{Mean field equation for CLVMD}
For the CLVMD, we set $\gamma_1=\gamma_2=\gamma$ and obtain the following reaction-diffusion equations:
\begin{eqnarray}
\frac{d c_A}{dt} &=& \mu c_A   + \gamma c_A (c_A +c_B)  - \gamma s c_A c_B  + D_A \dxx{c_A}, \\
\frac{d c_B}{dt} &=& \mu c_B   + \gamma c_B (c_A +c_B) + \gamma s c_A c_B + D_B \dxx{c_B}.
\end{eqnarray}
Again, from the above equations we get the following equations for the evolution of $c$ and $f$:
\begin{eqnarray}
\frac{d c}{dt} &=& \mu c \left( 1 - \frac{\gamma}{\mu}  c\right) + (D_A-D_B) \dxx{(c f)}+ D_B \dxx{c}, \\
\frac{d f}{dt} &=& \gamma s f (1-f) + D_A \dxx{f} + \frac{2D_A}{c} \dx{c}\dx{f} \nonumber \\ &+& f \frac{D_A-D_B}{c}\dxx{c} - f \frac{D_A-D_B}{c}\dxx{(cf)}.
\label{MFQ_CLVMD}
\end{eqnarray}
The homogeneous stable solution for the concentration equation is $c=\mu/\gamma$. Around this homogenous initial state, it is easy to see that the equations for CLVMD  and VMD model are identical.

We focus on the fixation probability $\mathcal{F}$ which is the likelihood that a certain species would prevail over the other, and ask for the dispersal strategy which would maximize $\mathcal{F}$. Here is a summary of our main results (i) In the VMD where the total number of individuals is strictly conserved, $\mathcal{F}$ is invariant with respect to change in diffusivity irrespective of selective advantage. Thus in this case the variation of dispersal rate is ineffective as a strategy. (ii) In the FVMD, birth and death lead to fluctuations of the total number of individuals. In the neutral case ($s=0$), $\mathcal{F}$ remains independent of diffusivity. However, a new effect arises when $s \neq 0$: the optimal strategy for disadvantaged individuals is then to move \emph{slower}. (iii) In the CLVMD, number fluctuations due to intra-species competition dominate; the best strategy to enhance $\mathcal{F}$ is then to move \emph{faster}. In a nutshell, dispersal strategies become crucial in the presence of number fluctuations; but exactly what the optimal strategy is depends on the form of non-conservation. Below we present details of our investigation which lead to these conclusions and also remark on the fixation times for each species in the neutral case.

An important point is that the mean-field descriptions of the three models are similar, as shown in Eqs. \ref{MFQ_VMD_FVMD}, \ref{MFQ_CLVMD}. Therefore for investigating the non-trivial effects that we highlighted in the introduction, we performed agent-based simulations for the models, supplement these by analytic arguments to support the numerical results.

%The space-time plots in Fig. \ref{World_Lines} show that the effect of population fluctuations and dispersal on the three models is very different. Interestingly, this happens even though their mean-field descriptions are similar (see Supplementary Material). 

\section{Methodology} 
We simulate the reaction along with the diffusion dynamics, by splitting the two processes. A single time step is broken into many substeps, each of duration $\Delta t$. At the first substep, for reactions, we implement the event-based variant of Gillespie algorithm \cite{Gillespie2007_1} at each site up to a small time $\Delta t$ and repeat it for all sites. After a reaction process, the spatial positions of individuals are updated for time $\Delta t$ according to their diffusivities. We follow the processes until global fixation is achieved. 

\subsection{Numerical Simulation}
In our numerical simulations, we start with $N/2$ individuals of type $A$ and an equal number of type $B$, placed randomly on a $1D$ lattice with $L$ sites with $N>L$. 
Thus the average density of individuals $\rho_0 = N/L > 1$. In each of the three models of population dynamics under study, there are two physical processes (i) reactions, namely birth, death, intra- and inter-species competition, and (ii) movement of individuals via diffusion. In our lattice model, reactions are on-site processes and diffusive moves happen between nearest-neighbour lattice sites. We split the reaction and diffusion processes, i.e., only reactions occur for a time interval $\Delta t$, followed by only diffusive moves for time $\Delta t$. The time interval $\Delta t$ is chosen to be much smaller than one Monte-Carlo step yet large enough that many reactions occur in $\Delta t$.

For reactions, we implement the event-based Gillespie algorithm \cite{Gillespie2007_1} at each site. To do so, we first calculate the propensity of all possible reactions on that site. A particular reaction would occur with a probability that is proportional to its rate. Let us consider a site $i$ containing $n_A$ individuals of type $A$ and $n_B$ number of type $B$. At this site, the total number of possible reactions due to birth, death, intra-species and inter-species reactions is
\begin{eqnarray}
\mathcal{R} &=& \left(\mu^{A}_{birth} + \mu^{A}_{death}\right) \, n_A + \left(\mu^{B}_{birth} + \mu^{B}_{death} \right) \, n_B + \gamma_1 \nonumber \\ &\times&  \left[n_A \left(n_A-1\right) + \, n_B \left(n_B-1\right) \right]  + 2 \left(\lambda + \gamma_2 \right) \, n_A \, n_B. 
\end{eqnarray}
We choose $\mu^{A}_{birth}=\mu^{B}_{birth}=\mu^{A}_{death}=\mu^{B}_{death}=\mu$ in our numerical simulation. The probability of a particular reaction event occurring is the ratio of the rate of the event to the total rate $\mathcal{R}$. For instance, $A$ would become $2A$ with probability $\frac{\mu \, n_A}{\mathcal{R}}$. If the reaction happens, the next step is to increase the time by $\ln(1/r)/\mathcal{R}$ where $r \in (0, 1]$ is a random number chosen from uniform distribution. The number of $A$'s and $B$'s change after each successive reaction, thereby affecting the total rate $\mathcal{R}$ at the site. Quite a few reactions occur until the sum of the reaction time-steps just crosses the chosen $\Delta t$. We follow the same procedure for all other sites. Recall that during this time step $\Delta t$, we solely do the reaction, and inter-site diffusive hops are not allowed.

Once the reactions are completed on every site of the system up to time $\Delta t$, we move a randomly chosen set of $ND \Delta t$ individuals to one of the nearest neighbor sites, where $D$ is the hopping rate of the species. The time evolution is averaged over a number $N_{hist}$ of histories, each starting from a new initial condition, a typical value of $N_{hist}$ being $30,000$.

\REM{%
 The values of the parameters that we have used for numerical simulations of the different models are given in Table I.

\begin{table}[!h]
\begin{center}
\scalebox{1}{%
 \begin{tabular}{c  c  c  c } 
 \hline
\multicolumn{1}{m{2.5cm}}{}  & VMD & FVMD & CLVMD \\
\hline
%\multicolumn{1}{m{2.5cm}}{Reactions:} & \multicolumn{1}{m{1.5cm}}{Eq.\,($1$)} & \multicolumn{1}{m{1.5cm}}{Eq.\,($2$)} & \multicolumn{1}{m{1.5cm}}{Eq.\,($3$)} \\
%\hline
%\multicolumn{1}{m{2.5cm}}{Reactions:} & $ A+B \xrightarrow {\lambda} 2A/2B$ [Eq.\,($1$)] & \multicolumn{1}{m{4.3cm}}{ $A+B \xrightarrow {\lambda} 2A/2B$; $A(\text{or} B)\xrightarrow {\mu}2A(\text{or} 2B)$; $A(\text{or} B)\xrightarrow {\mu} 0 $ [Eq.\,($2$)]} & \multicolumn{1}{m{6cm}}{$A(\text{or} B) \xrightarrow {\mu}2A (\text{or} 2B)$; $A+A\xrightarrow {\gamma_1} A$;  $B+B\xrightarrow {\gamma_1} B$;  $A+B\xrightarrow {\gamma_2} A (\text{or} B)$ [Eq.\,($3$)]} \\
%\vspace{0.8cm}
$\mu$ & $-$ & $2$ & $4$ \\ 
%\vspace{0.5cm}
$\lambda$ & $2$ & $2$ & $-$ \\ 
%\vspace{0.5cm}
$\gamma_1=\gamma_2 $ &  $-$ & $-$ & $1/8$ \\ 
%\vspace{0.5cm}
%Syatem size: $L$  & 128 & 128 & 128 \\
%\vspace{0.5cm}
%Initial density: $\rho_0$ & 64 & 64 & 64 \\
%\hline
%\vspace{0.5cm}
%Time step: $\Delta t$ & 1/64 & 1/64 & 1/64 \\
$D_A$ & 1 & 1 & 1 \\
$D_B$ & $0$ & $0$, $0.25$ & $0.96$, $0.98$ \\
\hline
\end{tabular}
}
\caption{\label{parameters} Parameter values for the three models. In most cases, we took $L=128$, $\rho_0=64$, and $\Delta t=\frac{1}{64}$. } \label{Parameters}
\end{center}
\end{table}
}

\section{Results}

In the following, let the diffusivities of the $A$ and $B$ species be $D_A = D$, and $D_B = D-\Delta D$ respectively.  In all our simulations we set $D=1$. 
We assume $\Delta D > 0$, this corresponds to $A$ particles diffusing faster than $B$ particles. We are interested in the effect of $\Delta D$ on the behavior of the models defined in Eqs.\, \eqref{Reaction_VM}$-$\eqref{Reaction_CLVM}, both when the selective advantage $s$ is zero (neutral case) and when it is nonzero. We take up these cases separately, focusing on the fixation probabilities $\mathcal{F}_A (s, \Delta D)$ and $\mathcal{F}_B (s, \Delta D)$, where the explicit dependence on $s$ and $\Delta D$ may be suppressed if not required. Evidently we have $\mathcal{F}_A+\mathcal{F}_B=1$. For the neutral case ($s=0$), we also discuss the fixation times $t_A$ and $t_B$ required to reach a state with all $A$ (all $B$). Mean fixation times $\langle t_A \rangle$ and $\langle t_B \rangle$ are quite different even when the fixation probabilities of both species are equal, a reflection of differences of dispersal dynamics when $\Delta D$ is nonzero. Below we present a detailed investigation  for the three models. We first discuss the case with $s=0$.

\subsection{Neutral Case ($s=0$)} 
\begin{figure}[ht!!]
\begin{minipage}{0.4\textwidth}
\includegraphics[width=\textwidth,height=0.22\textheight]{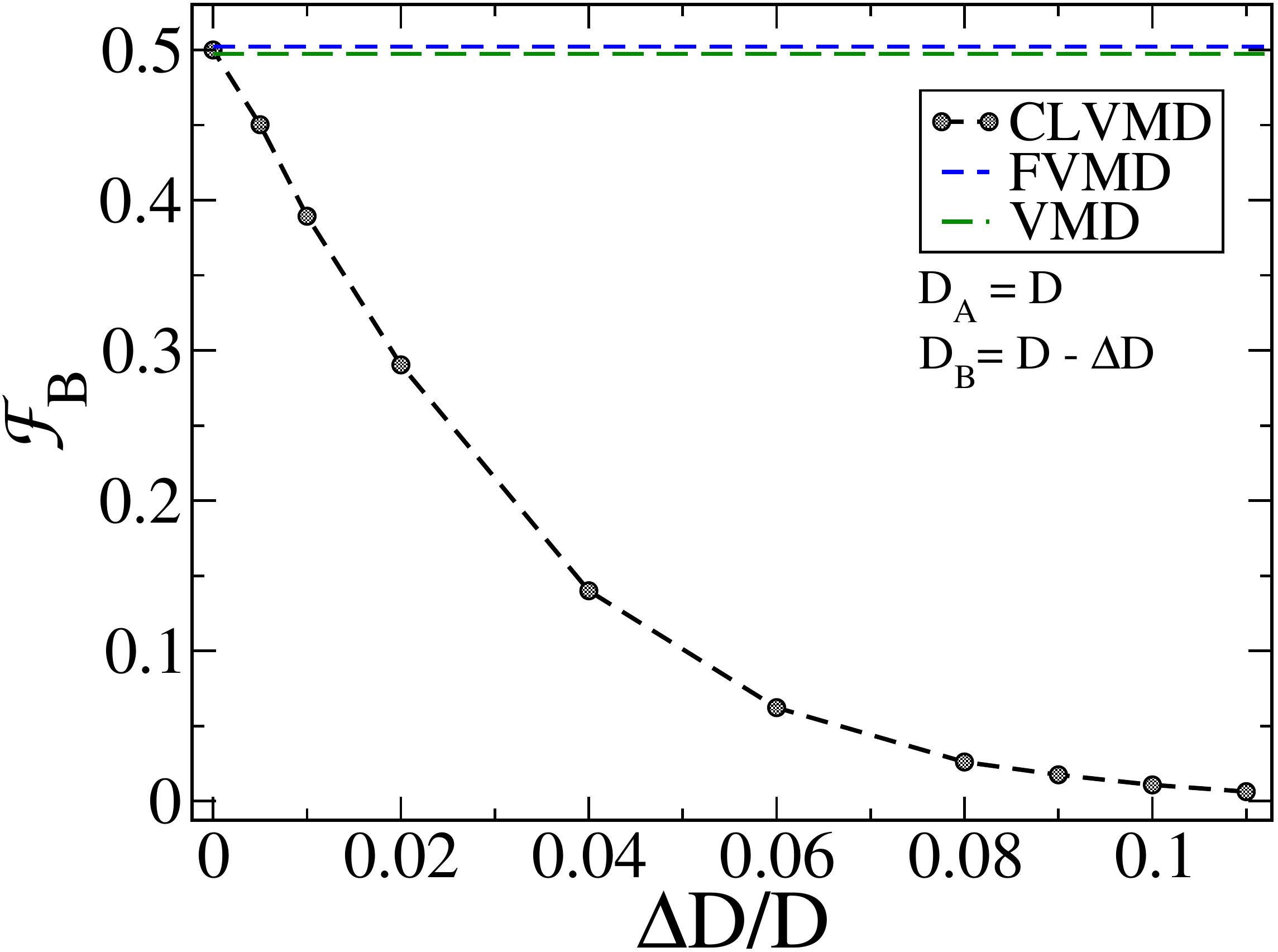}
\end{minipage}%
\caption{\label{Fixa_RelativeDiff} In the neutral case $s=0$, the fixation probability $\mathcal{F}_B$ does not vary with respect to relative diffusivity $\Delta D$ for VMD and FVMD, while it shows a strong variation with $\Delta D$ for the CLVMD.}
\end{figure}

Numerical simulations show that the variation of $\mathcal{F}_B(0, \Delta D)$ with $\Delta D$ is quite different for the three models (Fig.\, \ref{Fixa_RelativeDiff}). While $\mathcal{F}_B$ is immune to changing $\Delta D$ for the VMD and FVMD, it is very sensitive to $\Delta D$ for the CLVMD. These pronounced differences occur even though the mean-field equations for the concentration of species are the same for all three models.

\begin{table}
\scalebox{1}{
 \begin{tabular}{  l  l  l  l  l  l  l }
    \hline 
     {$L$ } & 
     \multicolumn{1}{m{2.5cm}}{\hspace{0.5cm}VMD} &
      \multicolumn{1}{m{2.5cm}}{\hspace{0.5cm}FVMD}  &
       \multicolumn{1}{m{2.5cm}}{\hspace{0.5cm}CLVMD} \\
\hline
   & $\langle t_A \rangle$ \hspace{0.5cm} $\langle t_B \rangle$ & $\langle t_A \rangle$ \hspace{0.5cm} $\langle t_B \rangle$ & $\langle t_A \rangle$ \hspace{0.5cm} $\langle t_B \rangle$ \\
    \hline
    32  & 146 \hspace{0.5cm} 955 & 118 \hspace{0.5cm} 232 & 172 \hspace{0.5cm} 110 \\
 %\hline
    64 & 587 \hspace{0.5cm} 4102 & 380 \hspace{0.5cm} 625 & 742 \hspace{0.5cm} 735 \\
%\hline
    128 & 2328 \hspace{0.4cm} 17721 & 1101 \hspace{0.4cm} 1587 & 1904 \hspace{0.4cm} 1841 \\
%\hline
  \end{tabular}
}
\caption{\label{FixationTime_table} Fixation time with unequal diffusivities for different system sizes $L$ where $\rho\equiv N/L=64$. We consider $D_B=0$ for the VMD and FVMD and $D_B=0.96$ for CLVMD.}
\end{table}

%%\begin{figure}[ht!!]
%%\begin{minipage}{0.4\textwidth}
%%\includegraphics[width=\textwidth,height=0.22\textheight]{All_FixationTimeUnEq.pdf}
%%\end{minipage}%
%%\caption{\label{FVMD_FixationTime} \textcolor{blue}{Mean fixation times $\langle t_A%\rangle$ and $\langle t_B\rangle$ grow with system size $L$ as $L^\theta$. In the range studied, $\langle t_A \rangle < \langle t_B\rangle$ holds for VMD and FVMD, while the reverse is true for CLVMD. We observe $\theta=2$ for the VMD with a conserved number of particles, and $\theta$ in the range $1.3$$-$$1.5$ for the FVMD and CLVMD, when the total number of particles is not conserved.}}
%%\end{figure}

\begin{figure*}[!ht]
\begin{minipage}{0.9\textwidth}
\includegraphics[width=\textwidth,height=0.195\textheight]{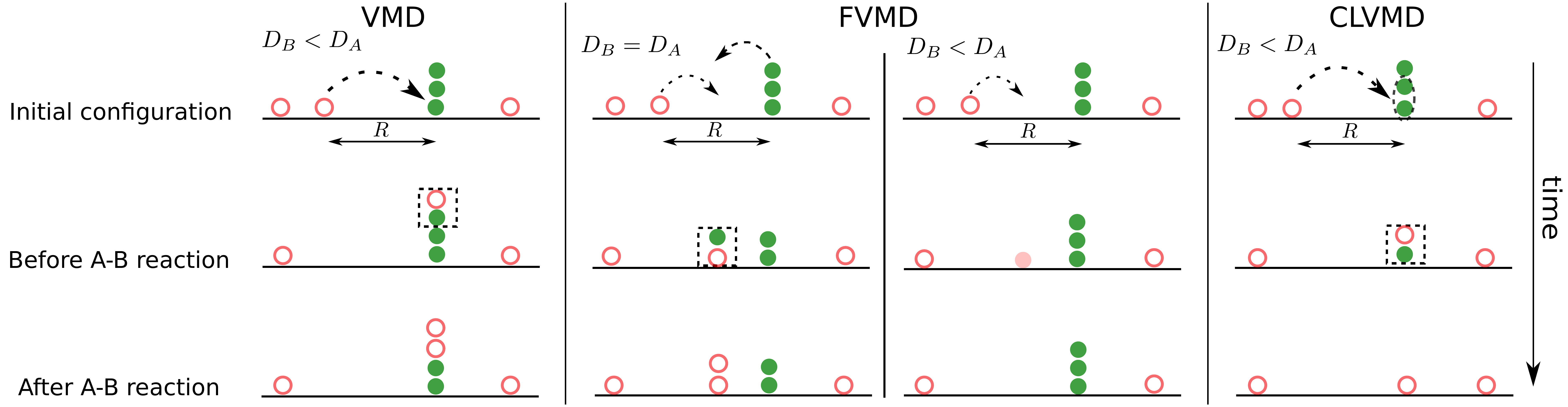}
\end{minipage}%
\caption{\label{Caricature} Cartoon of some moves and reactions to illustrate the contrasting effects of $\Delta D > 0$ in the three models when $s > 0$. Starting from the same initial configuration (top row) we depict configurations which occur just before (second row) and just after (third row) a possible $A$$-$$B$ interaction event. $A$ particles (open red circles) are stronger and faster; $B$ particles (solid green circles) are weaker and slower. VMD: Each reaction event can be mapped onto a step in the Moran process irrespective of diffusion constants. FVMD: In this model, an individual can die on its own. When the lifetime of an $A$ particle is less than $\tau_{Diff}$, it  is likely to die before interaction (depicted by the faint red circle in the second row), which is  advantageous for the weaker $B$ species, as the number of interactions would drop. Thus the optimal strategy for the weaker species is to make $\tau_{Diff}$ large, which it can do by moving as slowly as possible. CLVMD: The lifetime of a cluster of slow $B$'s formed by chance on a site is larger than that of  a corresponding $A$ particle cluster, leading to more intra-species extinction, in turn making it easier to fixate $A$.}
\end{figure*}

\emph{VMD}: The full dynamics involves diffusion with unequal $A$ and $B$ diffusivities along with $A$$-$$B$ symmetric reaction kinetics. The result of each reaction step is $2A$ or $2B$ with equal probability, implying that the overall numbers of $A$ and $B$ particles follow a Moran process. Thus $\mathcal{F}_A (0, \Delta D) = \mathcal{F}_B (0, \Delta D) =1/2$.

To understand the fixation dynamics, let us consider the case where $D_A=D$ and $D_B=0$.  Starting from uniformly distributed $A$ and $B$ particles, very quickly isolated pillars with a large number of $B$-particles are formed (see  Fig.~\ref{World_Lines}(a) and (b) (bottom-left)).  Consequently, the $A$ particle concentration in the vicinity of the pillar is strongly depleted.  Furthermore, because of number fluctuations, with finite probability the local pillar can eventually convert quickly into all $A$'s. This leads  to an increased local concentration of  $A$,  which then diffuse away as in Fig.~\ref{World_Lines}(bottom-left).   We have verified that this effect persists even when $D_B$ is nonzero, so long as $\Delta D > 0$.  

We observe that the mean fixation times for the species satisfy $\langle t_A \rangle < \langle t_B \rangle$ (see \autoref{FixationTime_table}).  First consider the case where $B$ fixates. At long times, we observe a single $B$ pillar with few scattered $A$ particles (see Fig.\,\ref{World_Lines}(b)).  The scenario is exactly opposite for the case of $A$-fixation, where the last $B$ pillar is in a sea of a  large number of $A$ particles (see Fig.~\ref{World_Lines}). Thus the frequency of reaction in the latter case is larger than for $B$-fixation. 

It is possible to obtain analytic results for the fixation probability as well as mean fixation times $\langle t_A \rangle$ and $\langle t_B \rangle$ in the limit of VMD dynamics in which one species (say $B$) has zero diffusivity, and the reactions happen infinitely fast. Consider the neutral case with an initial condition with $N_0=N/2$ $B$'s at a single site $S_0$ and no $A$ particles. At each time step, a single $A$ particle is assumed to reach $S_0$, on-site reactions being completed before the next $A$ arrives. We now show that the fixation probability $=1/2$ for both species, while mean times $\langle t_A\rangle$ and $\langle t_B\rangle$ for $A$ and $B$ fixation satisfy $\langle t_A\rangle < \langle t_B\rangle$.

Let us compute the probability $P_A(n)$ that global $A$ fixation occurs at the $n$'th step. $A$ particles arriving earlier at site $S_0$ must have converted to $B$, so that the number of $B$ particles after the $(n-1)$th step is $N_0+n-1$. The probability $Q(n-1)$ of this event is $g_0 \times g_1 \cdot \cdot \cdot  \times g_{n-1}$ where $g_i = (N_0+i)/(N_0+i+1)$ is the probability that an $A$ particle arriving at the $i$’th step is converted to $B$. Thus $Q(n-1)= N_0/(N_0+n-1)$. Now consider the arrival of the next $A$ particle. The probability that reactions result in the $N_0+n$ particles becoming all $A$'s is then $1/(N_0+n)$. If this happens, the system would  fixate globally to all $A$'s. Thus the probability $P_A(n)$ that $A$ fixation happens at the $n$’th time step is $Q(n-1) \times 1/(N_0+n)$. The overall $A$ fixation probability is then $\sum_{n=1}^{N_0} P_A(n)=1/2$, and the mean time of $A$ fixation is straightforwardly found to be 
\begin{equation}
\langle t_A\rangle = 2 N_0 \left [ H(2 N_0) -H(N_0) \right ] - (N_0-1)
\end{equation} 
where $H(M)= \sum^{M}_{m=1}  \frac{1}{m}$. For large $N_0$, it then follows that $\langle t_A\rangle  \approx C N_0$ with $C=(2 \, ln 2-1) \approx 0.386 N_0$.

On the other hand, in order for $B$ fixation to occur, the $B$'s must have survived at each of the earlier steps. The corresponding probability is $Q(N_0) = \frac{1}{2}$, and the corresponding survival time is $\langle t_B\rangle = N_0$. Evidently, $\langle t_A\rangle < \langle t_B\rangle$ holds. This result accords with the qualitative point that the faster species fixates earlier on average.

\emph{FVMD}: All reaction steps, whether interspecies competition or individual births or deaths, continue to maintain $A$$-$$B$ symmetry even if $\Delta D \neq 0$, as for the VMD. Thus inclusion of birth-death fluctuations in the FVMD thus does not change the conclusion $\mathcal{F}_A (0,\Delta D) = \mathcal{F}_B (0, \Delta D) =1/2$.  For the same reason as VMD, the inequality $\langle t_A \rangle < \langle t_B \rangle$ (see \autoref{FixationTime_table}) continues to hold for the FVMD.

\emph{CLVMD}: An advantage for the faster species results from a 
combination of unequal diffusivities and intra-species death terms (Eq.\,\eqref{Reaction_CLVM}), which  
act on several particles of the same species on the same site. Consider first the $A$ particles. 
Owing to the faster $A$ diffusivity, $A$-particle concentration fluctuations spread out quickly, 
thus minimizing the effects of intra-species death. Dispersal of $B$ particles is slower, 
so fluctuations which lead to $B$ particle clustering decay relatively slowly. 
This makes $B$ particles more prone to intra-species death. A depleted $B$ population on a given
site is then easier to convert to all $A$'s on that site through the 
competitive $A$ $-$ $B$ conversion terms (see Eq.~\eqref{Reaction_CLVM}) \cite{pigo13}. Thus overall, 
the optimal strategy to maximize the fixation probability within the CLVMD is to move fast.

We observe that  $\langle t_A \rangle > \langle t_B \rangle$ holds for relatively small $L$, but the difference narrows down as $L$ increases (see \autoref{FixationTime_table}). In the minority of cases in which the $B$ species does achieve fixation, it must be before the intra-species terms have had much effect. Thus $B$ fixation occurs at relatively early times.

\subsection{Role of Selective Advantage ($s>0$)} Going beyond the neutral case, in order to study the change of fixation probability brought in by unequal diffusivity, we define 
\begin{equation} 
\Delta \mathcal{F}_{B} (s, \Delta D) \equiv \mathcal{F}_{B} (s, \Delta D) -\mathcal{F}_{B} (s, \Delta D=0)
\label{Def:DeltaF}
\end{equation}
with $\Delta \mathcal{F}_{A}$ being defined similarly. Evidently we have $\Delta \mathcal{F}_A+\Delta \mathcal{F}_B=0$. Below, we see how $\Delta \mathcal{F}_{A,B}$ behaves in the three models under study.

\emph{VMD}: The argument used in the neutral case applies also when $s\neq 0$, implying that the fixation probability is the same as that of the Moran process with the corresponding value of $s$. The outcome of a particular Moran step does not depend on the manner in which an $A$ and a $B$ reach the same site (Fig. \ref{Caricature}). Hence $\Delta \mathcal{F}_A= \Delta \mathcal{F}_B=0.$

%Fixation times: \textcolor{blue}{To be added.}

\begin{figure}[H]
%\hspace{-0.95cm}
\centering
\begin{minipage}{0.41\textwidth}
\includegraphics[width=\textwidth,height=0.22\textheight]{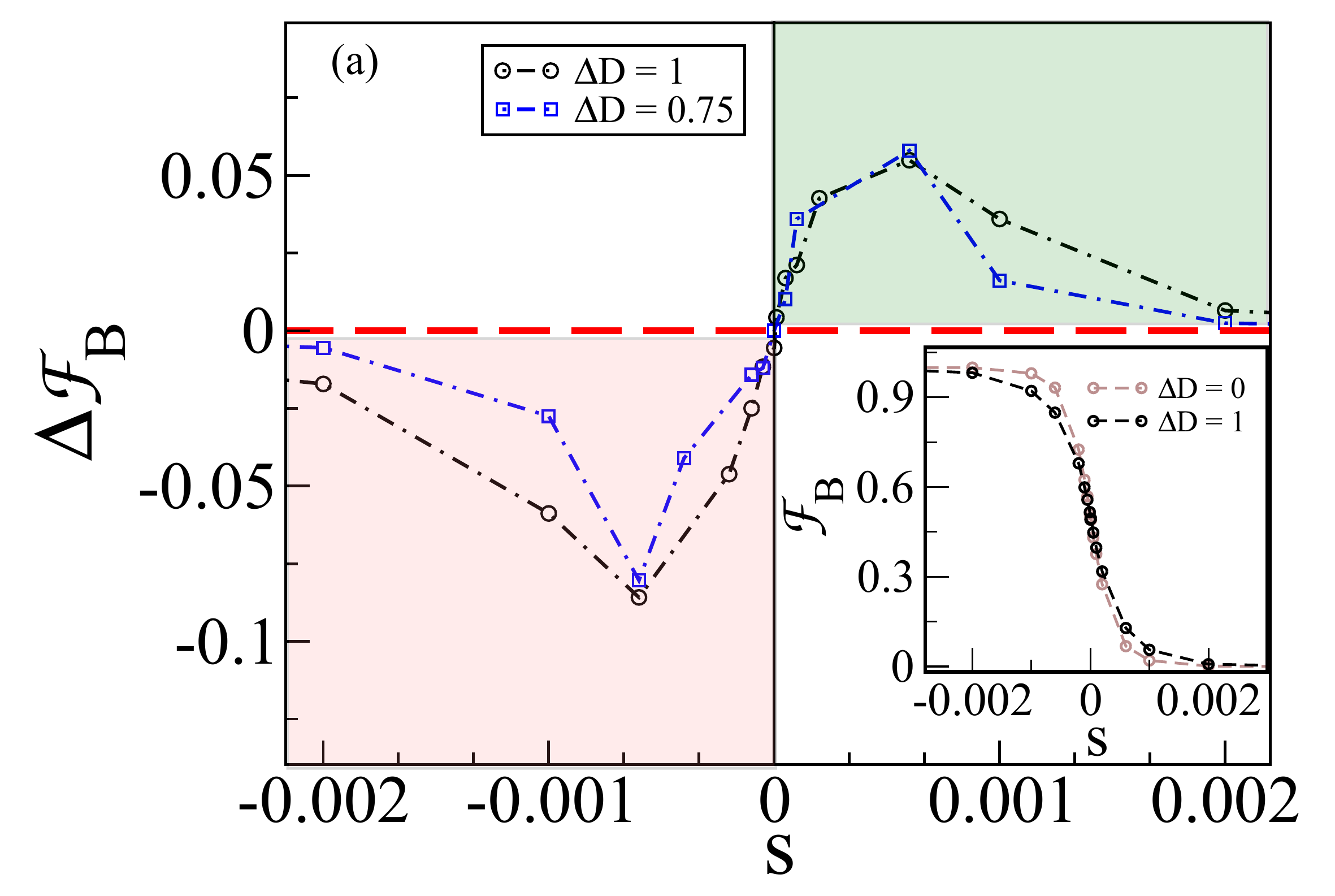} \\
\hspace{5cm} 
\includegraphics[width=0.95\textwidth,height=0.22\textheight]{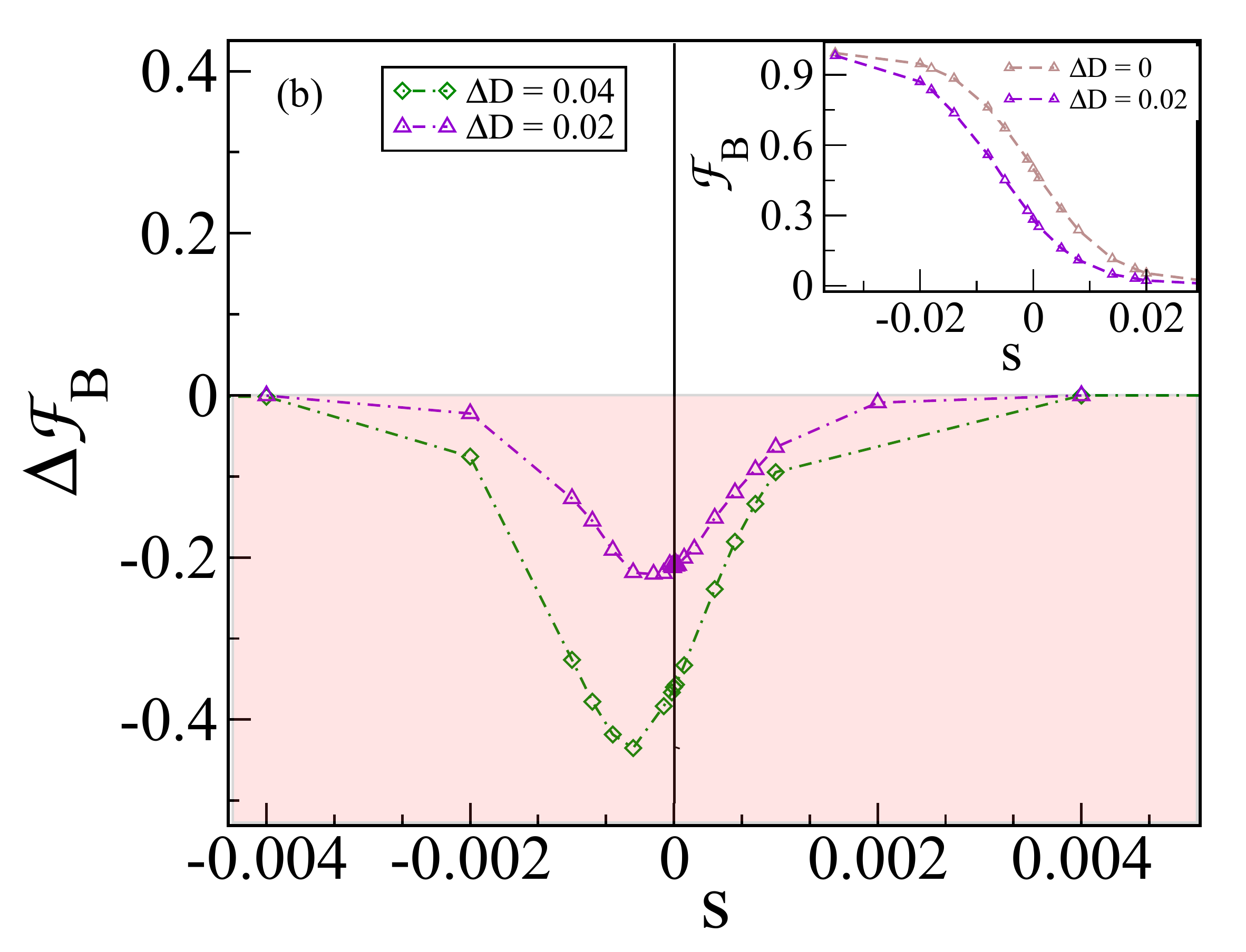} \\
\end{minipage}%
%\vspace{-1cm}
\caption{\label{Fixation_Diffusivity} Strategy plot: The change of $\Delta \mathcal{F}_B(s,\Delta D)$ of the slower $B$ species for given values of selective advantage $s$ for the three models; recall that  $s>0\,(s<0)$ implies selective advantage for $A(B)$.  VMD: $\Delta \mathcal{F}_B$ vanishes for all $s$, as depicted by the thick red line in (a). Diffusing faster or slower does not change the fixation probability. FVMD: When $B$ is weaker, $\Delta \mathcal{F}_B$ is positive implying that moving slowly is a better strategy than moving fast. $\Delta \mathcal{F}_B$ reverses sign when $s<0$, implying reversal of the strategy. CLVMD: the faster species gets a benefit for all values of $s$ as shown in (b). }
\end{figure}

\begin{figure}[!h]
\begin{center}
\begin{minipage}{0.4\textwidth}
\includegraphics[width=\textwidth,height=0.22\textheight]{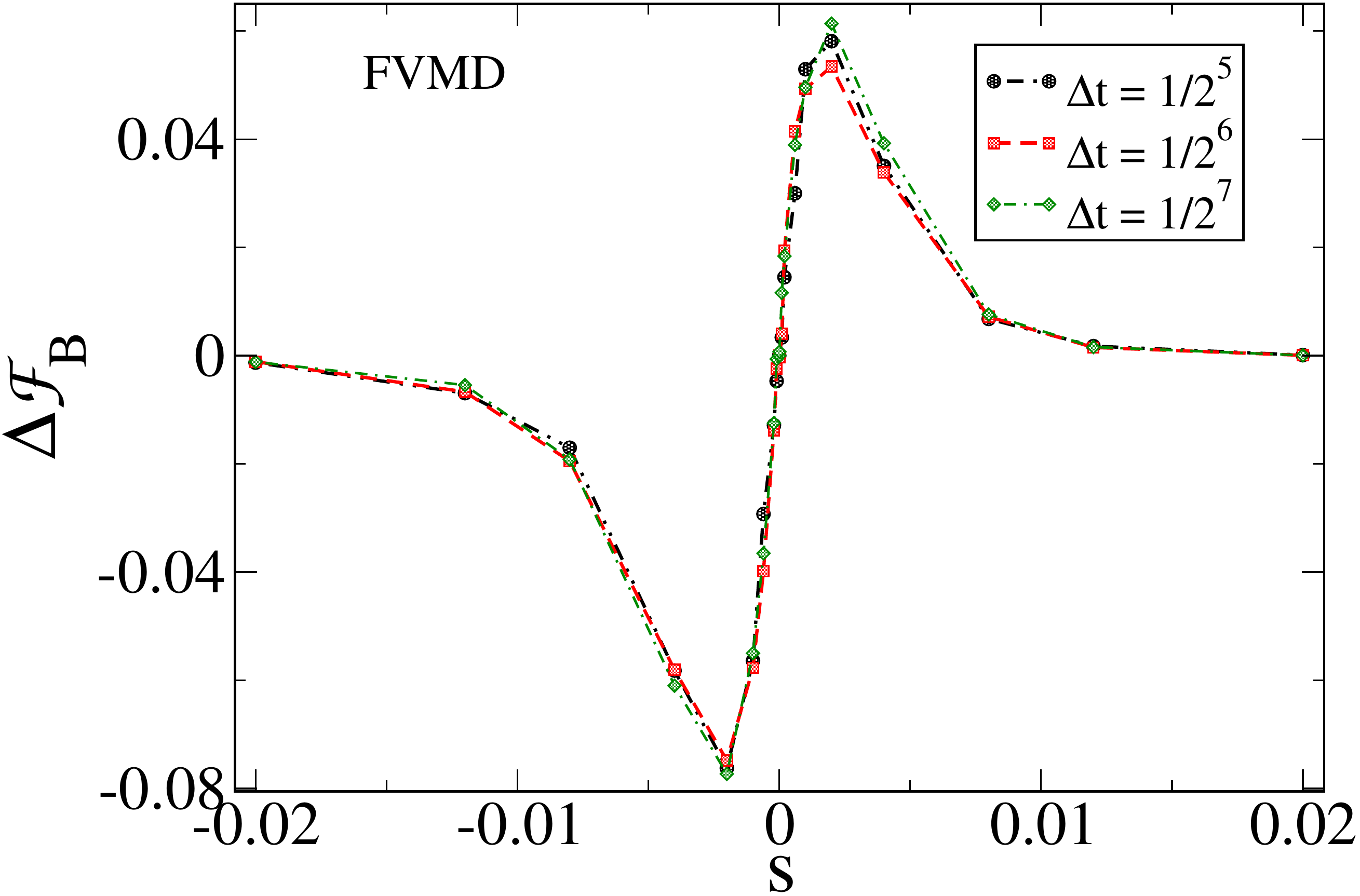}
\end{minipage}
\end{center}
\caption{\label{FVM_Fixa_Deltat} Plot of $\Delta\mathcal{F}_B(s,\Delta D)$ with respect to $s$ for different $\Delta t$ and with $\Delta D=1$  fixed.}
\end{figure}

\emph{FVMD}: The inclusion of birth-death processes has a strong effect and the best dispersal strategy now depends on the strength of the species. For instance, if $s>0$, then moving slowly is a better strategy than moving fast for the weaker $B$ species. Evidence for this comes from Fig.\,\ref{Fixation_Diffusivity} which shows that $\Delta \mathcal{F}_B$ is positive when $\Delta D$ and $s$ are both positive. To see how birth-death fluctuations can affect fixation, refer to Fig.\,\ref{Caricature} which compares the sequence of events with $D_B = D_A$ and $D_B < D_A$. The separation $R$ between an $A$ and $B$ particle undergoes a random walk
with an effective diffusion constant $D_{eff} = D_A + D_B = 2 D-\Delta D$. Thus, provided they survive, the typical time for the particles to meet is $\tau_{diff} \approx R^2/D_{eff}$. Further, owing to the birth-death process, the typical time of survival of an $A$ particle is $1/\mu$, implying there is a probability $(1-e^{-\mu \, \tau_{diff}})$ that the $A$ particle would not survive for time $\tau_{diff}$. Thus the best strategy for the weaker particle $B$ to survive is to increase $\tau_{diff}$ (decrease $D_{eff}$) to the maximum extent possible, which it can do by setting $D_B=0$, i.e., by standing still. This is corroborated by  Fig.\, \ref{Fixation_Diffusivity}, which shows that $\Delta \mathcal{F}_B$ is positive for $s>0$ meaning that the disadvantage for the $B$ species is reduced. The accrued advantage increases with $s$ for small $s$.  As $s$ increases in magnitude, selective advantage effects override the effects of unequal diffusivities, hence $\Delta {\mathcal F}_B \to 0$ as $|s|$ becomes large. This leads to the extrema in Fig.\,\ref{Fixation_Diffusivity}(a). In addition, we checked that our results do not depend on $\Delta t$ by varying a factor $4$,  and found that there is no appreciable change in our numerical estimates of the fixation probability (Fig.\,\ref{FVM_Fixa_Deltat}). 

\emph{CLVMD}: Reduction of $B$ diffusivity increases the residence time thus allowing more intra-species competition ($B+B\rightarrow B$) in time $\tau_{diff}$. The fall of the  number of $B$'s makes $A$ fixation more likely on that site. For $s=0$, we already showed the faster species is favoured, implying $\Delta \mathcal{F}_B < 0$.  The fact that $\Delta \mathcal{F}_B(s, \Delta D)\to0$ in Fig.~\ref{Fixation_Diffusivity}(b) as $|s|$ increases follows from the same considerations as in the FVMD.

%%%%%%%%%%%%%%%%%%%%%%%%%%%%%%%%%%%%%%%%%%%%%%%%%%%%%%%%%%%%%%%%%%%%%%%%%%%%%%%%%%%%
\REM{
\begin{figure}[ht!]
\begin{minipage}{0.44\textwidth}
\includegraphics[width=0.8\textwidth,height=0.18\textheight]{FVMDLambda0p001}
\end{minipage}%
\caption{\label{FM_SmallLambda} Time-sliced $\emph{f}\,(t)$ for the faster $A$ species in the FVMD with and without nonlinear intra-species competitive interactions. The curve with black dots indicates that $A$ fixation occurs preferentially at shorter times even though $\mathcal{F}(A)=\mathcal{F}_B=1/2$. A very small competitive nonlinear term changes both $\mathcal{F}$ and $\emph{f}\,(t)$.}
\end{figure}

To understand fixation dynamics in the neutral ($s=0$) case in systems with unequal diffusion constants, it
is instructive to study the \emph{Time-sliced fixation probability} $f(t)$, defined as the fractional number of $A$-fixations 
which occur at time $t$, i.e. $f (t) = m_A(t)=[m_A(t)+m_B(t)]$ where $m_A(t)$ $(m_B(t))$ is the number of histories
in which $A$ $(B)$ fixations occur at time $t$. For the neutral case with $D_A > D_B$, we find that although the overall
fixation probability $\mathcal{F}_A(0, \Delta D) = \mathcal{F}_B(0, \Delta D)= \frac{1}{2}$ for the VMD and FVMD, the corresponding time-sliced probabilities are not equal. Figure 6 shows that $A$ fixations dominate ($f(t) > \frac{1}{2}$)  at relatively short
times, while $B$ fixations dominate at later times. This result is in consonance with $\langle t_A \rangle < \langle t_B\rangle$  and the explanation thereof, as discussed earlier. In the CLVMD, even a very small logistic nonlinearity has a
strong effect on $f(t)$ (Fig. 6). At small times, before the nonlinearity kicks in, the variation is very similar to
that of the FVMD; but at longer times $f (t)$ shows a nonmonotonic rise as the preference for the faster
species is manifest, consistent with $\mathcal{F}_A(0, \Delta D) > \frac{1}{2}$.}

\section{Conclusions}
In conclusion, our results for three well known models of competing populations with unequal diffusivity 
show that it is crucial to account for fluctuations beyond mean field theory to understand their behavior
and formulate dispersal strategies. The fixation probability is maximized by increasing the dispersal rate if
intra-species competition is present; but in a situation where the species is disadvantaged and subject to
fluctuations due to birth and death, fixation probability is maximized by moving slowly; and it is unaffected
by dispersal if competing interactions are described by number-conserving dynamics that can be mapped
onto a Moran process. It would be interesting to explore the effects of fluctuations in broader contexts,
such as competing populations in compressible flows, or with clustered initial conditions. Finally, in a
large population, it may be interesting to ask for an optimal strategy to achieve a larger fixation probability
within a fixed time. Our studies of mean fixation times \REM{and time-sliced fixation probabilities} constitute a
step in this direction.

\section{Acknowledgments}
The authors acknowledge support from  intramural funds at TIFR Hyderabad from the
Department of Atomic Energy (DAE). MB acknowledges support under the DAE 
Homi Bhabha Chair Professorship of the Department of Atomic Energy.

\section{Appendix}
\subsection{Appendix A: Quasi-stationary state and global extinction in FVMD}
\begin{figure}[!h]
\begin{center}
\includegraphics[width=0.75\linewidth, height=0.6\linewidth]{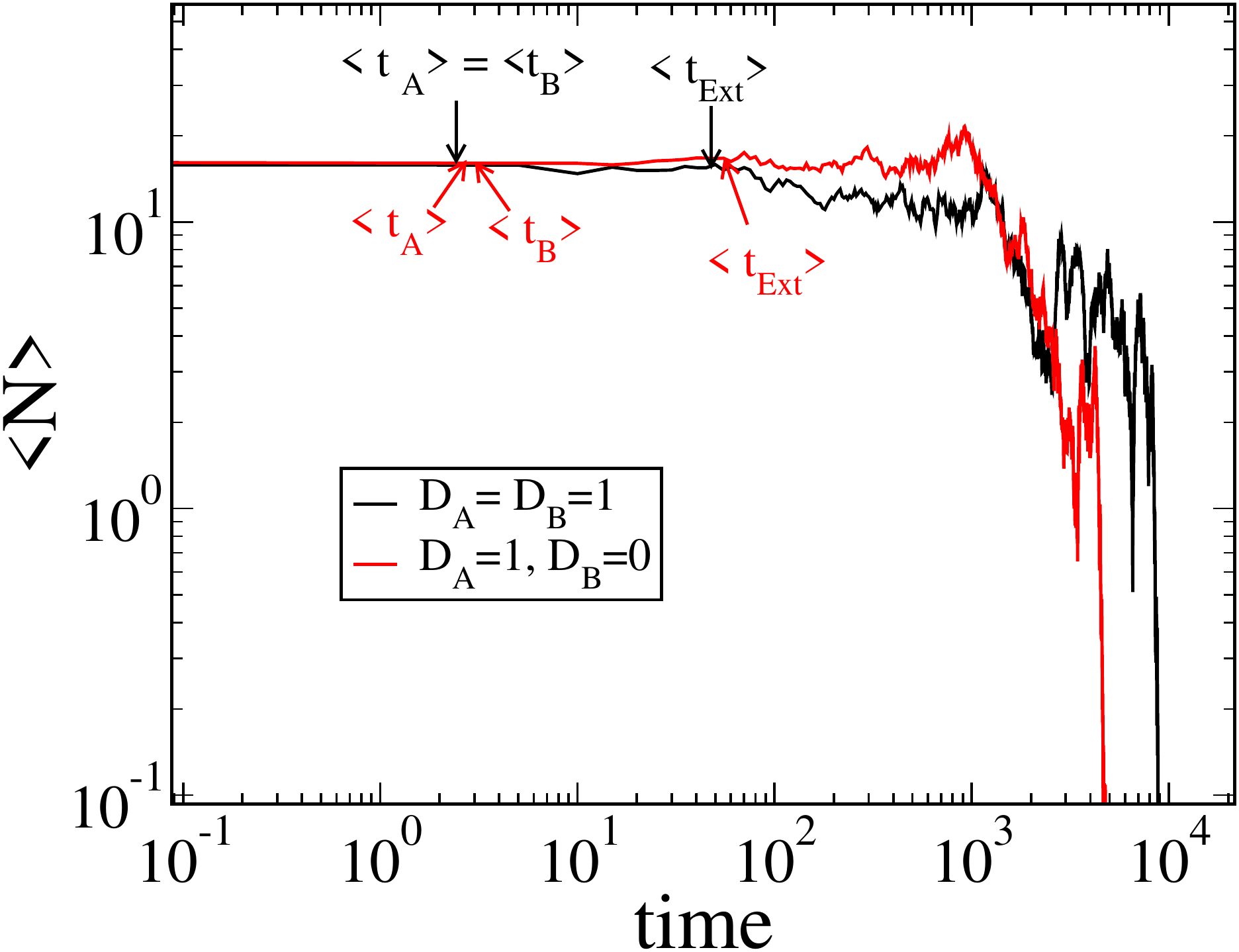}
\vspace{1cm}
\includegraphics[width=0.75\linewidth, height=0.6\linewidth]{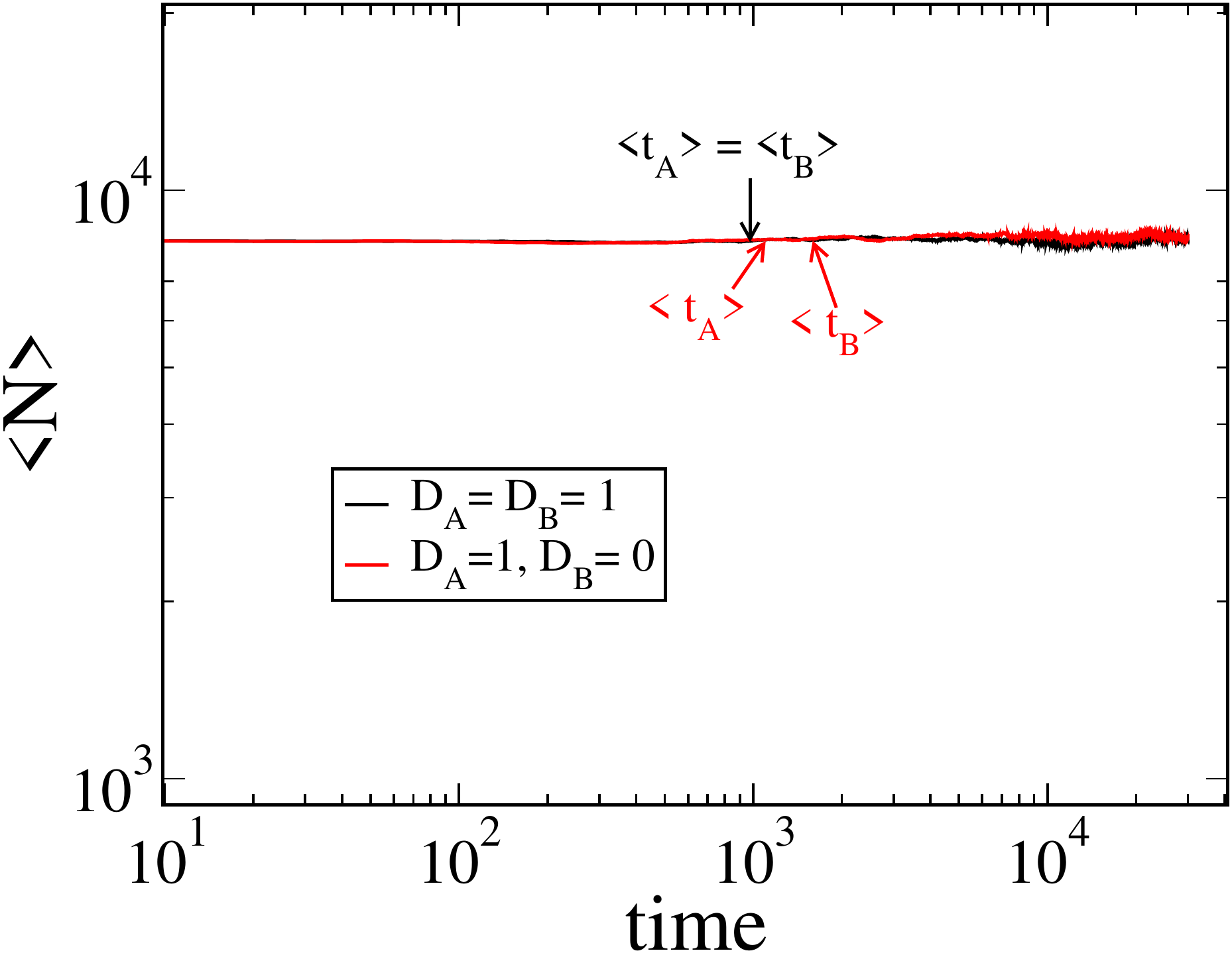}
\end{center}
\caption{\label{Fixation_FVMD_L128} (top) Long time dynamics showing extinction for our simulation with $L=8$ and $\rho=2$.  (bottom) Average number of particles $\langle N(t) \rangle$ versus time $t$ for equal (black solid line) and unequal (red solid line) diffusivities.  The angular brackets denote averaging over $10000$ independent realizations. Average fixation ($\langle t_A \rangle$, $\langle t_B \rangle$) and extinction ($\langle t_{Ext} \rangle$) times are marked with arrows. }
\end{figure}

In this appendix we show the presence of a long lived quasi-steady state (QSS) in which $\langle N \rangle$ fluctuates around a constant value before eventual extinction. Our numerical simulations, even with small system size $L=8$ and density $\rho=2$, demonstrate that the fixation of one of the two species occurs well before extinction [see Fig. \ref{Fixation_FVMD_L128}(top)]. For the large system sizes used in the main text, we are always in the QSS regime [see Fig. \ref{Fixation_FVMD_L128}(bottom)].

\REM{%
In order to examine, how the average total number $\langle N (t)\rangle$ changes around the average fixation and beyond that, we show $\langle N(t) \rangle$ with equal and unequal diffusivities as shown in \autoref{Fixation_FVMD_L128}(left). We choose the same values of parameters namely, system size $L=128$ and density $\rho=64$ as we have considered for our results in the main text. For these choices of parameters, extinction occurs at very long times which is beyond the scope of this work. 

However, for much smaller system, namely $L=8$ and  low density $\rho=2$, we demonstrate how $\langle N(t) \rangle$ varies with time and how far the average fixation is located from the average extinction times, we mark $\langle t_A \rangle$, $\langle t_B \rangle$, and $\langle t_{Ext} \rangle$  by arrows in \autoref{Fixation_FVMD_L128}(right) for the case with equal ($D_A=D_B=1$) and unequal diffusivities ($D_A=1$, and $D_B=0$). This shows that the fixation occurs at very early times compared to the extinction, and $\langle N(t)\rangle$ fluctuates around the initial number near the far beyond from the fixation.  %
}

\REM{
\begin{figure}[!h]
\begin{center}

\end{center}
\caption{\label{Fixation_FVMD_L8} \textcolor{blue}{For FVMD, $\langle N (t) \rangle$ with $t$ is plotted with equal (black solid line) and unequal diffusivities (red solid line). For FVMD, extinction occours at very large time whereas fixation occurs a long before than the extinction. When $D_A=D_B$, average fixation time $\langle t \rangle$ is marked by the black-arrow, and we show $\langle t_A \rangle$ and $\langle t_B \rangle$ by red-arrows while the diffusivities are unequal.}}
\end{figure}
}

\end{document}